\pdfoutput=1
\documentclass[aps,prl,a4,10pt,twocolumn]{revtex4-2}

\usepackage{amsmath,amssymb}
\usepackage{mathtools}
\usepackage{graphicx}% Include figure files
\usepackage{dcolumn}% Align table columns on decimal point
\usepackage{bm}% bold math
\usepackage{graphicx}
\usepackage{float}
\usepackage{subcaption}
\usepackage{physics}
\usepackage{xcolor}
\usepackage{hyperref}% add hypertext capabilities
\usepackage{color}
\definecolor{Blue}{rgb}{0.00, 0.00, 0.80}
\definecolor{Red}{rgb}{0.80, 0.00, 0.00}
\definecolor{Green}{rgb}{0.00, 0.50, 0.00}
\definecolor{Purp}{rgb}{0.50, 0.00, 0.50}
\newcommand{\black}{\color{black}}
\hypersetup{
    colorlinks=true,       % false: boxed links; true: colored links
    linkcolor=black,          % color of internal links (change box color with linkbordercolor)
    citecolor=blue,        % color of links to bibliography
    filecolor=magenta,      % color of file links
    urlcolor=blue           % color of external links
}

\begin{document}

\preprint{APS/123-QED}

\title{Stochastic resetting prevails over sharp restart for broad target distributions}

\author{Martin R. Evans}
\email{m.evans@ed.ac.uk}
\affiliation{SUPA, School of Physics and Astronomy, University of Edinburgh, Peter Guthrie Tait Road, Edinburgh EH9 3FD, UK}
\author{Somrita Ray}
\email{somrita@iiserbpr.ac.in}
\affiliation{Department of Chemical Sciences, IISER Berhampur, Odisha 760003, India}
\date{\today}% It is always \today, today,

\begin{abstract}
Resetting has been shown to reduce the completion time for a stochastic process, such as the first passage time for a diffusive searcher to find a target.
The time between two consecutive resetting events is drawn from a waiting time distribution $\psi(t)$, which defines the resetting protocol.
Previously, it has been shown that deterministic resetting process with  a constant time period, referred to as sharp restart, can minimize the mean first passage time to a fixed target.  Here we consider
the more realistic problem of a target positioned at a random  distance $R$ from the resetting site, selected from a given target distribution $P_T(R)$. We introduce the notion of a conjugate target distribution to a given waiting time distribution. The conjugate target distribution, $P_T^*(R)$,  is that $P_T(R)$ for  which $\psi(t)$ extremizes the mean time to locate the target. In the case of diffusion we derive an explicit expression for $P^*_T(R)$ conjugate to a given $\psi(t)$ which holds in arbitrary spatial dimension. Our results show that stochastic resetting prevails over sharp restart for  target distributions with exponential or heavier tails.
\end{abstract}

\maketitle

Resetting a stochastic process to begin anew  may drastically improve the time to complete a task, such as locating a target \cite{EMS20}. In a nutshell, resetting can cut off errant trajectories that take the process away from its desired end. The idea has proven useful in many different contexts including
optimizing the performance of %randomized 
computer algorithms \cite{LSZ93,BRH22,BRH24}, chemical reactions \cite{RUK14,RRU15,BPMR23}, animal foraging \cite{BS14,VCMLNA22,EMS22}, biophysical processes \cite{RLSG16,Brf2020a} and other searches \cite{CM15,BDBR16,PR17,CS18,B18,BMS23,SBEM24}.
In the last decade, diffusion with stochastic resetting has received particular attention \cite{EM11a,EM11b,WEM13,EM14,MSS15,CS15,KGN15,KT19,JBPD,
EM16,NG16,PKE16,G19,BS20a,BRR20,MMSC21,DBMS22,S23,DBM23,MOK23,SBEM23,BFM23}
as it provides a simple paradigm where analytic results may be obtained. 
%{\bf define diffusion with resetting}. 
It is well known that the mean time for a diffusive process to locate a fixed target, the mean first passage time (MFPT), diverges. Notably, the introduction of a resetting rate, at which the process is restarted from some chosen initial position, renders the MFPT finite. Moreover,
there is an optimal resetting rate which minimizes the MFPT \cite{EM11a}. 

\begin{figure}[t!]
 %\vspace*{5cm}
 \centering
    \includegraphics[width = 0.48\textwidth]{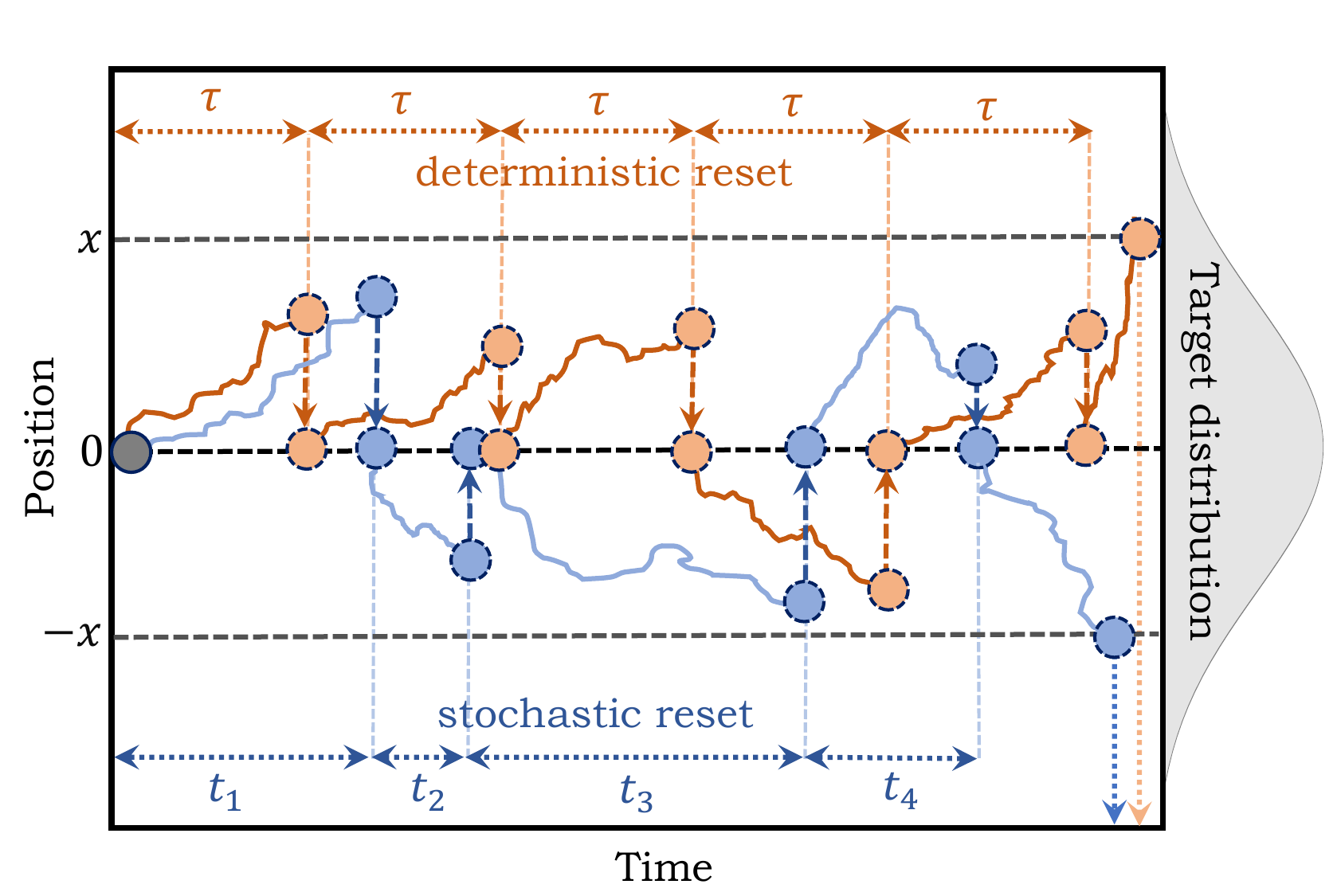}
    \caption{\small A schematic position vs. time diagram for diffusion with resetting (in one dimension). An absorbing target is located at a  random distance $R=|\vec x|$, chosen from a distribution of target centred around the origin (in grey). Two types of resetting protocols are illustrated. The red trajectory is subject to deterministic, periodic reset, i.e., where two consecutive resetting events occur after a constant interval of time $\tau$. The blue trajectory is subject to stochastic reset, i.e., where the time intervals $t_i$ between the $(i-1)^{\rm th}$ and $i^{\rm th}$ reset are taken from a waiting time distribution $\psi(t)$.\vspace{-0.4cm}} 
    \label{fig:model}
\end{figure}

A more general formulation of the resetting process is to consider, rather than a resetting rate,  the waiting time  distribution, $\psi(t)$, of the random time $t$ between resets.
This distribution defines the resetting strategy.
A constant resetting rate $r$, referred to as Poissonian resetting,  has  waiting time distribution $\psi(t) = r {\rm e}^{-rt}$ \cite{EM11a}, whereas resetting with a deterministic period $\tau$,  referred to as sharp restart or periodic resetting \cite{ER20,ER21}, has $\psi(t)= \delta(t-\tau)$ [see Fig.~\ref{fig:model}].
Both Poissonian and periodic resetting have been realized experimentally \cite{TPSRR20,BBPMC20,FBPCM21,BFPCM21}.
One can also consider other distributions $\psi(t)$ such as power-law distributions \cite{EM16,PKE16,NG16}.

For diffusion with resetting the choice of $\psi(t)$
%waiting time distribution 
that minimizes the MFPT is, in fact, sharp restart \cite{PKE16,BDBR16,PR17,CS18}, where the period $\tau$ has to be chosen optimally, given a fixed distance $R_0$ from the resetting position to the target.
This result was somewhat anticlimactic as it implied that stochasticity in the reset times was not advantageous.  However, the optimization of $\tau$ relies on the precise knowledge of the distance to the target, which is %somewhat 
unrealistic from a practical point of view.\\
\indent
A more realistic scenario  \cite{EM11b,KGN15,KT19}
is where the target is located at a {\it random} distance $R$ from an expected location, say the origin, [see Fig. \ref{fig:model}]. {\black The scenario we have in mind is best illustrated through the everyday example of searching for one's keys. Let us assume there's a location where our keys should be kept, their {\em home} let's say.  But usually  when we go to retrieve our keys they are not where they should be and are instead located at some random distance from their designated home. This distance will have a distribution which will be broader depending on how untidy we are. 
In an ecological context animals search for some target such as food, a mate or shelter within range of their home, to which they return repeatedly. This is known as home range search \cite{PKR20,ABGMRTRR24}. The target they seek will be situated at some random distance from the home, which is not a priori known.
More generally the distribution of targets is of importance in many search problems \cite{Bell,BLMV11} as it plays a crucial role in the efficacy of the search protocol \cite{TVB12}.
}\\
\indent
{\black The question we  consider is that of the optimal search strategy for a {\it general} target distribution, which has not been adequately addressed so far in the literature. As we shall see, the optimal search can be strikingly different depending on the nature of the target distribution. 
To elucidate the question of optimal strategy, we
define a  {\em conjugate target  distribution}, $P_T^*(R)$, to a given waiting time distribution, $\psi(t)$, as that target distribution for which $\psi(t)$ is optimal. That is, it extremises the MFPT averaged over the target distribution
\begin{equation}
    \overline{ \langle T_r \rangle} = \int_0^\infty{\rm d}R\, 
    % S_{d}(R)
     P_T(R)\langle T_r(R) \rangle\;,\label{Tav}
\end{equation}
where $\langle T_r(R) \rangle$ is the MFPT to a target at $R$ under resetting to the origin, with waiting time distribution $\psi(t)$. 
%{\black We will return to the question of whether the conjugate distribution is unique in the conclusion.}
For {\black an underlying diffusive search process}  we are able to obtain an explicit formula for the conjugate distribution.}
The significance of our result
is that it furnishes a relation between a resetting protocol, characterized by a waiting time distribution, and its optimal target distribution. This relation  will allow an informed choice of  resetting protocol  suitable for a given problem where one has a knowledge of $P_T(R)$.
In particular, our formulae  reveal that stochastic resetting protocols are optimal when the target distribution is broad, e.g. an exponential or a power law.
Thus, generally stochastic resetting prevails over sharp restart.\\
\indent
{\black To begin, we  assume a one-dimensional geometry and later generalize to $d>1$.}
The  target distance from the origin, $R$,  is drawn from a symmetric target distribution $P_T(R)$ \cite{EM11b} at the start of the search; it is then natural to  take the
origin as the resetting position.
{\black We emphasize that the random target position is fixed during the search and one has to perform the average \eqref{Tav} over the target position.
In contrast, if the position $R$ were redrawn after each reset  sharp restart would be optimal \cite{PR17}.} \\
%Thus the two problems are fundamentally distinct.\\
% and ask what is the waiting time distribution $\psi(t)$ that minimizes the expected time to locate the target. 
%How to optimize resetting for a given  target distribution is an open problem that we address in this work.
\indent
The explicit formula for the conjugate target distribution $P^*_T(R)$ that we obtain for diffusion with stochastic resetting  with waiting time distribution $\psi(t)$  reads
\begin{subequations}
\begin{eqnarray}
   P_T^*(R) &=& Z^{-1}f^2(R),\label{PT}\\
   \mbox{where}\;\;\; f(R)&=& \int_0^\infty {\rm d}t\, \psi(t)(1- \,Q_0(R, t))\,.\label{PT_f}
\end{eqnarray}
\end{subequations}
 Here $Z = \int_0^\infty {\rm d}R\,  f^2(R)$ is the normalization constant. {\black The utility of the formula contained in Eqs. (\ref{PT}, \ref{PT_f}) is that one simply evaluates the integral \eqref{PT_f}
for any given waiting time distribution $\psi(t)$ to obtain the conjugate target distribution.}
In Eq. \eqref{PT_f}, $Q_0(R,t)$ is the survival probability for free diffusion, in the absence of resetting, commencing at distance $R$ from an absorbing target. In $1d$, it
is well known that 
\begin{equation}Q_0= \mbox{erf} (R/\sqrt{4Dt})\;,\label{Q0}
\end{equation}
where $D$ is the diffusion constant \cite{Redner,BMS13}.
{\black In the following we evaluate \eqref{PT_f} for three important waiting time distributions: Poissonian, delta function (sharp restart) and power law.
}

We first check that Eq. \eqref{PT}, the conjugate target distribution to an arbitrary $\psi(t)$ for  one-dimensional free diffusion with resetting, is a proper distribution. 
Clearly it is positive and {\black it can  be shown \cite{ER24-SM} that the normalization $Z$ in Eq.~\eqref{PT}
is finite if $\psi(t)$ decays faster than $t^{-5/4}$ for large $t$, i.e. 
$\langle t^{1/4} \rangle =\int_0^\infty {\rm d}t\, \psi(t) t^{1/4}$ is finite.}
%
%$Z \leq \left( \frac{4D}{\pi} \right)^{1/2} %\langle t^{1/2} \rangle$, where }. Thus a %sufficient condition for  the conjugate of %$\psi(t)$, Eq. \eqref{PT}, to be normalizable is that the expectation value of $t^{1/2}$ is finite, where $t$ is the random waiting time.}

We now consider some particular choices of $\psi(t)$ and evaluate $P_T^*(R)$ in $d=1$ from Eq. \eqref{PT}.
First we take the case of Poissonian resetting $\psi(t) =
r {\rm e}^{-rt}$ and find
the conjugate target distribution to be,
 from
(\ref{PT},\ref{PT_f})
\begin{equation}
P_T^*(R) = \sqrt{\frac{r}{D}}{\rm e}^{-2 \sqrt{\frac{r}{D}} R},
\label{PTp}
\end{equation}
which itself is  an exponential distribution. With these choices of $\psi(t)$ and $P_T^*(R)$ one finds from Eq.~\eqref{Tav} that  $\overline{ \langle T_r \rangle} = 1/r$, which is finite for $r>0$. {\black Equation \eqref{PTp} previously appeared  in \cite{KGN15}, where a Laplace target distribution was proposed
on information theoretic grounds and  the MFPT was optimized for Poissonian resetting. Our result goes further to show that  for a Laplace target distribution, Poissonian resetting is optimal out of all resetting strategies.} 

%%%%%%%%%%%%%%%%%%%%%%%%%%%%%%%%%%%%%%%%%%%%%%%%%%
\begin{figure*}[ht!]
 \centering
     \includegraphics[width = 0.345\textwidth]{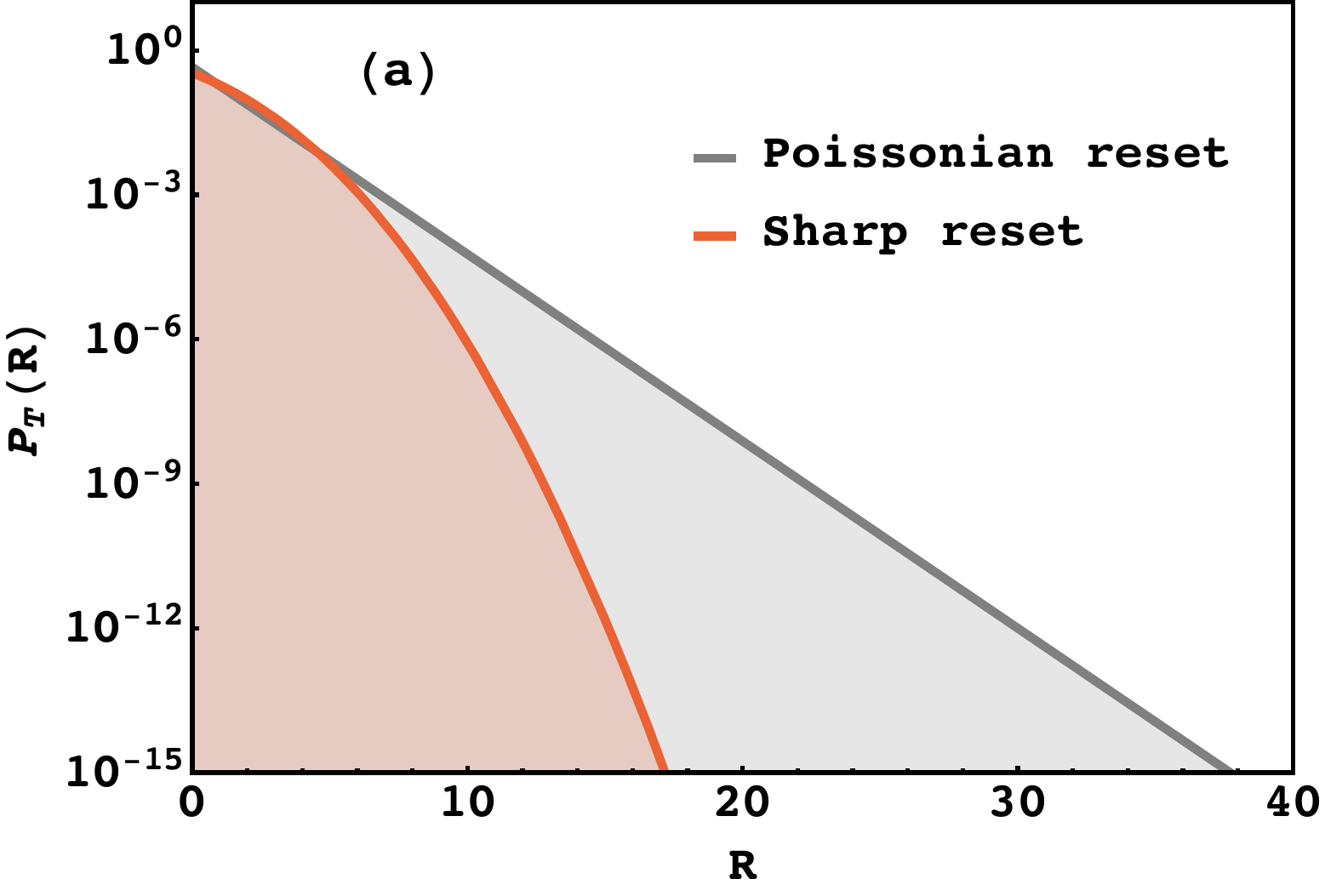}
    \includegraphics[width = 0.322\textwidth]{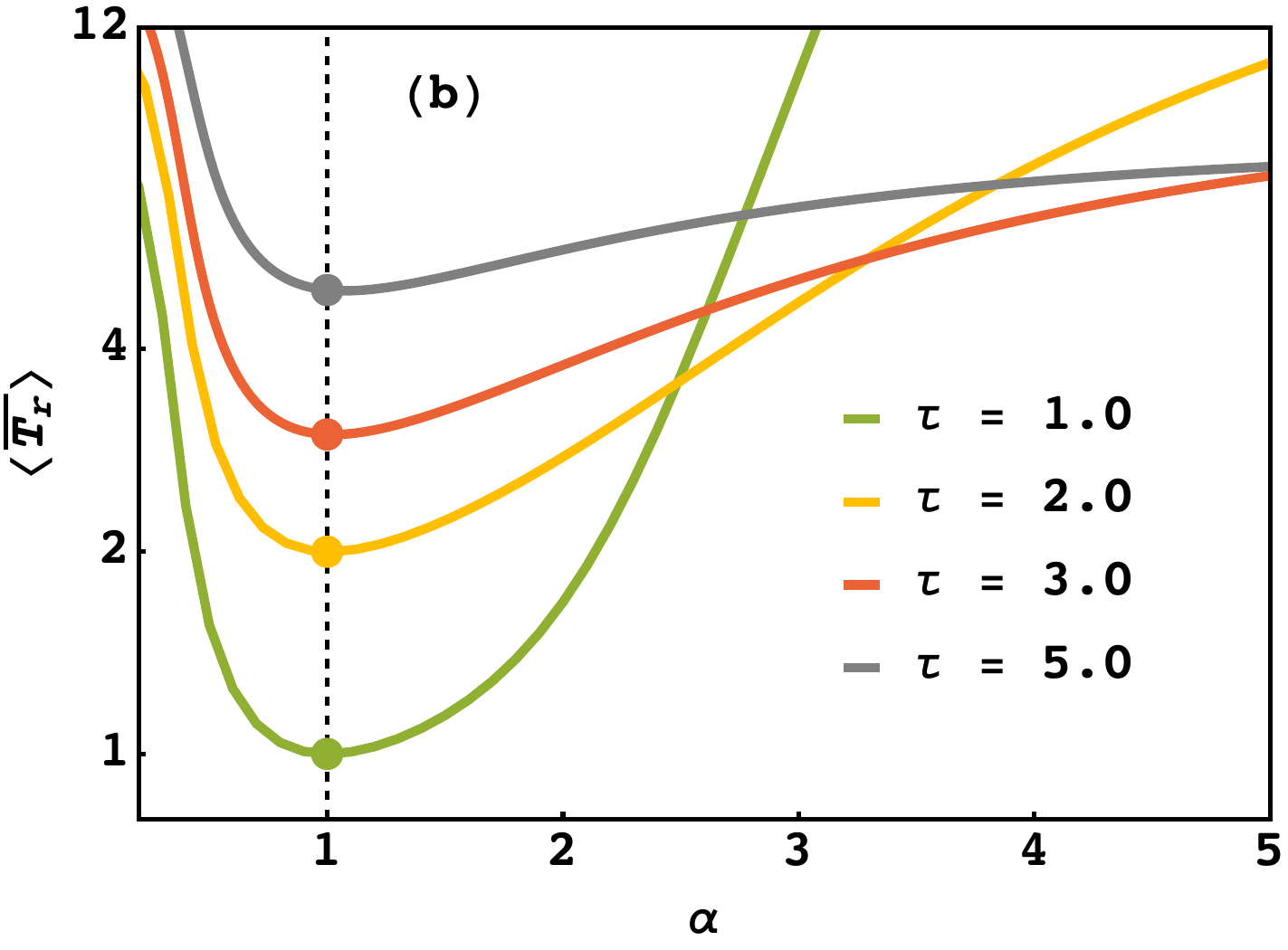}
    \includegraphics[width = 0.32\textwidth]{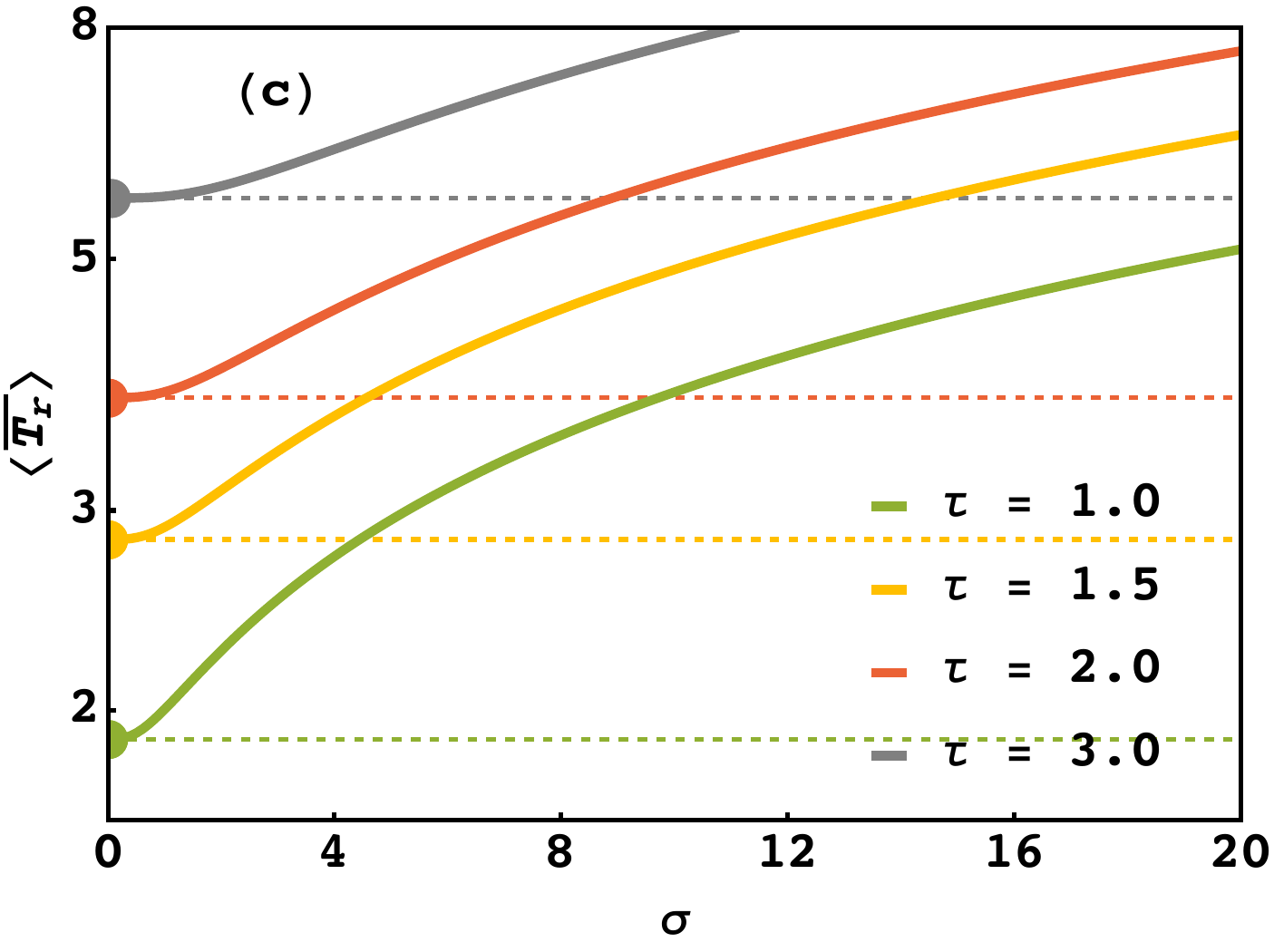}
    \caption{\small {\bf (a)} Target distributions $P_T^*(R)$ conjugate to waiting time distributions $\psi(t)$ [see Eqs.\eqref{PT} and \eqref{PT_f}].
     Grey curve:  Poissonian resetting [from Eq. \eqref{PTp}, with $\tau = 1/r = 5$]. Red curve: sharp restart [from Eq. \eqref{PTs}, with $\tau = 5$]. For the same $\tau$, the distribution conjugate to  Poissonian resetting has a heavier tail compared to that of sharp restart. {\bf (b)} MFPT averaged over $P_{T}^*(R)$ [from Eq. \eqref{PTp}], obtained from a general (normalized) waiting time distribution $\psi_1(t)$ [Eq. \eqref{psi1}]
    that reduces to Poissonian reset for $\alpha=1$. {\bf (c)} MFPT averaged over $P_{T}^*(R)$ [from Eq. \eqref{PTs}], obtained from a general (normalized) waiting time distribution $\psi_2(t)$ [Eq. \eqref{psi2}] that reduces to sharp reset when $\sigma\to 0$. 
    In each case shown in panels (b) and (c), the minimum $\overline{ \langle T_r \rangle}$ (obtained for $\sigma\to 0$ and $\alpha=1$, respectively) is marked by colored discs, which confirms our result. For all cases, $D=1$.  }
    \label{fig:schem}
   \end{figure*}
%%%%%%%%%%%%%%%%%%%%%%%%%%%%%%%%%%%%%%%%%%%%%%%%%%
%%%%%%%%%%%%%%%%%%%%%%%%

Next, we consider 
%the conjugate target distribution for 
sharp restart 
%\cite{PKE16,BDBR16,PR17,ER20,ER21} 
where 
$\psi(t) = \delta(t-\tau)$ and $\tau$ is the deterministic reset period.
The conjugate target distribution from Eq. (\ref{PT},\ref{PT_f}) is
\begin{equation}
P_T^*(R) = \,Z^{-1}  \mbox{erfc}^2( R/\sqrt{4D\tau})\;,
     \label{PTs}
\end{equation}
where $Z=(4 (2-\sqrt{2}) \sqrt{D \tau})/\sqrt{\pi}$. 
It is interesting to contrast  the distributions Eqs. \eqref{PTp} and \eqref{PTs} [see Fig.~\ref{fig:schem}(a)]. The first is a broad distribution with exponential decay length $\sqrt{D/4r}$ whereas the latter has a Gaussian tail $\sim {\rm e}^{-R^2/2D \tau}$ i.e. it is a narrow distribution with effective cutoff at 
$\sqrt{2D\tau}$. Indeed when sharp restart is used with target distribution Eq. \eqref{PTp}
one finds that
the integral in Eq. \eqref{Tav} for the MFPT always diverges for any $\tau$ \cite{KBGN17}.
This is because for sharp resetting, the MFPT for a point target at distance $R$
[given by Eq.~\eqref{Tavp}] diverges as ${\rm e}^{R^2/4D\tau}$ for large $R$ and fixed $\tau$, and this divergence dominates when the average Eq. \eqref{Tav}  is taken over a broad target distribution such as  Eq.~\eqref{PTp}.
 Thus sharp resetting is not a good strategy when the target has a broad distribution decaying more slowly than
$P_T(R) \sim {\rm e}^{-R^2/4D\tau}$ and in this case Poissonian resetting clearly prevails over sharp restart. On the other hand, when 
%the distribution of $R$ 
$P_T(R)$ has a tail that decays faster, as with Eq.~\eqref{PTs}, sharp restart will be the superior strategy.\\
\indent
As noted above, sharp reset has been of interest as  it minimizes the MFPT when the distance $R_0$ to the target is known \cite{CS18}.
This situation corresponds to a target distribution $P_T (R) = \delta(R-R_0)$, which evidently does not coincide with Eq. \eqref{PTs}.
The resolution is that  sharp restart is a particular case where
another extremum of the MFPT occurs, driven by the constraint of $\psi(t)$ being a positive distribution [see \cite{ER24-SM} for details]. {\black However, generally we expect the conjugate distribution $P_T^* (R)$
%Eq.~(\ref{PT},\ref{PT_f}) 
to be unique.}

Another relevant choice of $\psi(t)$ is a heavy-tailed distribution. We choose $\psi(t) = \Gamma^{-1}(\alpha-1){\rm e}^{-1/t} t^{-\alpha}$ which is normalizable for $\alpha > 1$. Then $P_T^*(R)$ may be found from $f(R)$ [Eq. \eqref{PT_f}], computed using identity 4.3.9 of \cite{NG1969}
\begin{equation}
    f(R) = \frac{\Gamma(\alpha{-}\frac{1}{2})}{\pi^{1/2}\Gamma(\alpha)} \frac{1}{\tilde R^{p+1}} {}_2F_1(\alpha-1,\alpha-\frac{1}{2},\alpha; -\frac{1}{\tilde R^2})\;, \label{PTh}
\end{equation}
where $\tilde R = R/(4D)^{1/2}$, $p= 2\alpha -3$  and ${}_2F_1(a,b,c;z)$ is the standard hypergeometric function. The large $R$ behaviour is $P_T^*(R) \sim R^{-4(\alpha-1)}$. Thus the conjugate of a heavy-tailed waiting time distribution with exponent $\alpha$ is a heavy-tailed target distribution with exponent $4(\alpha-1)$.
 {\black Interestingly, the condition for the  MFPT  averaged over the conjugate target distribution Eq.~\eqref{Tav} to be finite is $\alpha >7/4$. This corresponds to a target distribution with finite variance, a condition that was also found in \cite{KT19}.}

We now give a derivation of the result 
Eq.~(\ref{PT_f},\ref{PT}). 
We begin with the general first renewal equation \cite{PKE16,CS18,EMS20}  for $Q_r(R,t)$, the survival probability in the presence of resetting, 
 where the initial position of the particle and  the resetting position are both taken as the origin and the absorbing target is located at distance $R>0$
\begin{multline}
Q_r(R,t) = \Psi(t) Q_0(R,t) \\
+ \int_0^t{\rm d}\tau\, \psi(\tau) Q_0(R,\tau)Q_r(R, t-\tau)\;.\label{surv}
\end{multline}
Here $\Psi(t) = \int_t^\infty {\rm d}\tau\, \psi(\tau)$ is the probability of no reset up to time $t$ and $Q_0(R,t)$ is the survival probability up to time $t$ in the absence of resetting. Thus the first term in Eq. \eqref{surv} is the probability of no resets and survival up to $t$. The second term in Eq. \eqref{surv} integrates the probability of first reset at time $\tau$ and survival: the integrand is the probability of first reset at time $\tau$, survival up to time $\tau$ with no resetting, then survival from $\tau$ to $t$ with resetting. 

Eq. \eqref{surv} may be solved by Laplace transform to yield
\begin{equation}
     \widetilde Q_r(R,s)  = \frac{ \int_0^\infty {\rm d}t\,{\rm e}^{-st} \Psi(t) Q_0(R, t)}{1- \int_0^\infty {\rm d}t\,{\rm e}^{-st} \psi(t)\,Q_0(R, t)},
     \label{surv_LT}
\end{equation}
where the Laplace transform is defined as $\widetilde Q_r(R,s) = \int_0^\infty {\rm d} t\,{\rm e}^{-st}Q_r(R, t)$.
The MFPT is obtained by setting $s=0$, and integration by parts gives, see \cite{PKE16,CS18,ER24-SM}
\begin{equation}
    \langle T_r(R)  \rangle = \frac{ \int_0^\infty {\rm d}t\, \psi(t) \int_0^t{\rm d}t'\,Q_0(R, t')}{\int_0^\infty {\rm d}t\, \psi(t)(1- \,Q_0(R, t))}\;.\label{Tavp}
\end{equation}
We now average the MFPT over the target distribution $P_T(R)$, where $R$ is  the distance of the target from the resetting site,
and denote the average as $\overline{ \langle T_r \rangle}$
[Eq.~\eqref{Tav}].
%\begin{equation}
%    \overline{ \langle T_r \rangle} = \int_0^\infty{\rm d}R\, 
    % S_{d}(R)
%     P_T(R)\langle T_r(R) \rangle\;.\label{Tav}
%\end{equation}
In order to extremize Eq.~\eqref{Tav} with respect to $\psi(t)$ we take the functional derivative [See \cite{ER24-SM} for derivation]
\begin{multline}
    \frac{\delta \overline{\langle T_r \rangle}}{\delta \psi(t')}
=\mbox{\hspace{-0.1cm}}\int_0^\infty\mbox{\hspace{-0.35cm}}{\rm d}R\, P_T(R)\left[
    \frac{ \int_0^{t'} {\rm d}\tau\,Q_0(R, \tau)}{\int_0^\infty {\rm d}t\, \psi(t)(1- \,Q_0(R, t))} \right.  \\
    - \left. \frac{ (1-Q_0(R,t'))\int_0^\infty {\rm d}t\, \psi(t) \int_0^t{\rm d}\tau\,Q_0(R, \tau)}{\left[\int_0^\infty {\rm d}t\, \psi(t)(1- \,Q_0(R, t))\right]^2}
    \right]\;.\label{dTdpsi}
\end{multline}
Eq.~(\ref{dTdpsi}) holds for a general first passage process, with survival probability $Q_0(R,t)$.
%,  and a general waiting time distribution $\psi(t)$.
%Ideally, for a given $P_T(R)$,  we would like to find the optimal $\psi(t)$, for which the r.h.s of Eq. \eqref{dTdpsi} is zero for all $t'$. However this is a difficult, general problem. 
%We instead  consider the converse: 
%We now  seek to find the target distribution for which Eq. \eqref{dTdpsi} vanishes for all $t'$, for a {\em given} $\psi(t)$. 

The conjugate target distribution Eq.~\eqref{PT} is  that which renders a particular waiting time distribution optimal, in the sense that $\psi(t)$ extremizes the MFPT averaged over this target distribution.
{\black Thus we seek to find the target distribution for which Eq.~\eqref{dTdpsi} vanishes for all $t'$, for a given $\psi(t)$. Remarkably, for an underlying diffusive search, the ansatz Eq.~(\ref{PT},\ref{PT_f}) achieves this. }%We solve this problem for diffusion with stochastic resetting. 
%We show that in the case of diffusion with resetting the ansatz Eq.~\eqref{PT_f}
% renders $\psi(t)$ extremal---remarkably, it 
%ensures  the functional derivative Eq. \eqref{dTdpsi} is equal to  zero $\forall t'$. 
To show this we insert Eqs.~(\ref{PT},\ref{PT_f}) into Eq.~\eqref{dTdpsi} and require that
\begin{multline}
%Z\frac{\delta \overline{\langle T_r \rangle}}{\delta \psi(t')}=
\int_0^\infty\mbox{\hspace{-0.2cm}}{\rm d}R \left[\int_0^\infty \mbox{\hspace{-0.2cm}}{\rm d}t\, \psi(t)
      (1- \,Q_0(R, t))\int_0^{t'} \mbox{\hspace{-0.2cm}}{\rm d} \tau\, Q_0(R,\tau)\right.  \\ \left.
    - (1-Q_0(R,t'))\int_0^\infty\mbox{\hspace{-0.2cm}} {\rm d}t\, \psi(t) \int_0^t{\rm d}\tau\,Q_0(R, \tau) \right]=0. \label{cond}
\end{multline}
Taking the Laplace transform with respect to $t'$ with Laplace variable $s$ and after relabelling of integration variables we obtain the condition [See \cite{ER24-SM} for derivation]
\begin{multline}
\int_0^\infty {\rm d}t\int_0^\infty\mbox{\hspace{-0.2cm}} {\rm d}t' 
\left[ {\rm e}^{-st'} \psi(t)- {\rm e}^{-st} \psi(t')\right] I(t,t') =0,
\label{condlt}
\end{multline}
where
\begin{equation}
I(t,t')= \int_0^{t'}\mbox{\hspace{-0.2cm}}{\rm d}\tau\int_0^\infty \mbox{\hspace{-0.2cm}}{\rm d}R\,(1- \,Q_0(R, t))Q_0(R,\tau). \label{Idef}
\end{equation}
Note that  the term in the square bracket in Eq.~\eqref{condlt} is anti-symmetric in the integration variables $t,t'$. Therefore,
if  $I(t,t')$ is symmetric in $t,t'$, the integrand is antisymmetric under interchange of $t,t'$ and we conclude that the integral in Eq.~\eqref{condlt} is identically zero.

 For diffusive search where $Q_0(R,t)$ is given by Eq.~\eqref{Q0}, the integrals in  \eqref{Idef} are performed using identity 4.7.1  of \cite{NG1969}
\begin{equation}
    \int_0^\infty \mbox{\hspace{-0.3cm}}{\rm d} R\,\, \mbox{erfc}(aR) \mbox{erf}(bR)\,
    = \frac{1}{\sqrt{\pi}b}\left( \frac{\sqrt{(a^2+b^2)}}{a}-1\right)\label{erfident}
\end{equation}
to give
\begin{eqnarray}
    I(t,t') 
    &=&\frac{8}{3}\sqrt{\frac{D}{\pi}}\left[(t+t')^{\frac{3}{2}}-t^{\frac{3}{2}} - (t')^{\frac{3}{2}}\right]\;.
\end{eqnarray}
Thus, $I(t,t')$ is symmetric
under $t \leftrightarrow t'$, therefore  Eq. \eqref{condlt} holds, from which it follows that Eq. \eqref{cond} holds $\forall t'$.

Our derivation  shows that Eqs.~(\ref{PT}, \ref{PT_f}) are extremal { for a diffusive search but it  remains to confirm that $\psi(\tau)$  minimizes Eq.~\eqref{Tav} and hence is optimal. To do so,
we now consider a couple of specific cases numerically. 
First, to confirm our claim that the target distribution given in Eq. \eqref{PTp} is that for which 
%to the case of free diffusion under 
Poissonian resetting is optimal, we consider a general waiting time distribution 
\begin{equation}
\psi_{1}(t)=\mathcal{N}_1e^{-(t/\tau)^{\alpha}}\,, \label{psi1}
\end{equation} where $\mathcal{N}_1 = \alpha/[\tau\Gamma(1/\alpha)]$ is the normalization constant. Therefore, $\psi_1(t)$ reduces to Poissonian resetting  with rate $r=\tau^{-1}$ when $\alpha \to 1$. We use Eq.~\eqref{Tav} to numerically compute the MFPT averaged over $P_T^*(R)$ [taken from Eq. \eqref{PTp}] with $\psi_1(t)$ for different values of $\alpha$. We discover that $\overline{\langle T_r \rangle}$ is always minimal for $\alpha=1$ [see Fig.~\ref{fig:schem}(b)], thus confirming that Poissonian resetting  {\it minimizes} the MFPT computed with 
Eq. \eqref{PTp}, within the class of waiting time distributions given by Eq. \eqref{psi1}.

 Next, we consider the case of sharp restart. Taking a general Gaussian waiting time distribution 
\begin{equation}
 \psi_{2}(t)= \mathcal{N}_2 e^{-(t-\tau)^2/2\sigma^2}\, ,   \label{psi2}
\end{equation} where $ \mathcal{N}_2 = \sqrt{2/\pi} ({\sigma[1+\erf({\frac{\tau}{\sqrt{2}\sigma}})]})^{-1}$, we see that it reduces to sharp resetting with period $\tau$ as the standard deviation $\sigma \to 0$. Calculating $\overline{\langle T_r \rangle}$ numerically with $P_T^*(R)$ [taken from Eq.~\eqref{PTs}] and $\psi_2(t)$ as a function of $\sigma$ for different values of $\tau$, we observe that for each $\tau$, $\overline{\langle T_r \rangle}$ is minimum when $\sigma\to 0$ [see Fig.~\ref{fig:schem}(c)]. In other words, the MFPT averaged over the target distribution is minimized for the class of waiting time distributions given by Eq.~\eqref{psi2} when $\psi_{2}(t)$ reduces to $\psi(t)=\delta(t-\tau)$. These two specific cases thus suggest that for a  well-behaved waiting time distribution $\psi(t)$, there exists an optimal target distribution that {\it minimizes} $\overline{\langle T_r \rangle}$.

{\black Importantly, formulae Eq.~(\ref{PT},\ref{PT_f}) also hold in  dimension $d>1$,
for a target distribution that is symmetric about the
origin, so that there is no angular dependence.
One just needs to generalize the normalization in Eq.~\eqref{PT} to
 $Z = \int_0^\infty {\rm d}R\,S_d(R)  f^2(R)$ 
 where $S_{d}(R)=
    \frac{\pi^{d/2}}{\Gamma(d/2)} R^{d-1}$,
    so that  $\int_0^\infty {\rm d}R \;S_d(R)\, P_T^*(R)=1.$
    The proof of Eqs. (\ref{PT},\ref{PT_f}) can be  generalized to  $d>1$, where in order to have a target of finite extent, it is convenient to consider 
an absorbing $d$-dimensional sphere of radius $a$
centered at the target position $R$ \cite{EM14}.} %centred at  $\vec x$ (where $|\vec{x}|=x > a$) 
%The particle starts at the origin  and undergoes diffusion with diffusion constant $D$ and stochastic resetting to the origin. 
%When the particle reaches the surface of the target sphere, it is absorbed.
%For general $d$ the Laplace transform of the survival probability without resetting  is
%known in closed form 
%\begin{equation}
%    \tilde Q_0(R,s) = \frac{1}{s} \left[ 1 - \left(\frac{R}{a}\right)^\nu \frac{K_\nu(R(s/D)^{1/2}}{K_\nu(a(s/D)^{1/2}}\right]\label{survLT},
%\end{equation}%
%where $\nu = 1-d/2$ and $K_\nu(z)$ is a modified Bessel function of the second %kind \cite{Redner,BMS13}. 
%The proof of Eqs. \eqref{PT} and \eqref{PT_f} follows the one-dimensional case, but we now take the double Laplace transform
%$\tilde I(u,v) = \int_0^\infty {\rm d}t\,{\rm e}^{-ut}\int_0^\infty {\rm d}t'\,{\rm e}^{-vt'}I(t,t')$ 
%and require that it
%is symmetric under  $u \leftrightarrow v$.
%This can be shown  using  integral identities for  the %product of functions $K_\nu(z)$ \cite{GR}. 
%Thus in any  dimension $d$,
%Eq. \eqref{PT} gives the particular target %distribution for which an arbitrary $\psi(t)$ %is extremal. 
As an example, for  the Poissonian case,
 $\psi(t) = r {\rm e}^{-rt}$, %the integral in Eq.~\eqref{PT} can  be performed {\black for arbitrary $d$}  and 
 we obtain  for arbitrary $d$ 
\begin{equation}
    P_T^*(R) = Z^{-1} \left[ \left(\frac{R}{a}\right)^\nu \frac{K_\nu(R(r/D)^{1/2}}{K_\nu(a(r/D)^{1/2}}\right]^2\;,
    \label{PTd}
\end{equation}
where $\nu = 1-d/2$ and $K_\nu(z)$ is a modified Bessel function of the second \cite{ER24-SM}.
This distribution is normalizable for $d<4$.
In 1$d$ one can check using $K_{1/2}(z) = (\pi/2z)^{1/2} {\rm e}^{-z}$ that the target distribution reduces to Eq.~\eqref{PTp} above.
%P_T^*(R) = \alpha_0 {\rm e}^{-2 \alpha_0 R}$ with %$\alpha_0=\sqrt{r/D}$ 
%as found above.

In conclusion, we have defined the target distribution  conjugate to a waiting time distribution $\psi(t)$ as the target distribution for which  $\psi(t)$ extremizes the MFPT. In the case of diffusion with stochastic resetting we found a simple expression Eqs.~(\ref{PT}, \ref{PT_f}) for the conjugate target distribution. In $d=1$ we have shown that the target distribution conjugate to exponential waiting time, the case of Poissonian resetting, is itself an exponential distribution, whereas the conjugate distribution to sharp restart is given by Eq.~\eqref{PTs}, which decays more quickly for large $R$. Consequently, Poissonian resetting outperforms sharp restart when the target distribution decays more slowly than $P_T(R) \sim {\rm e}^{-R^2/4D\tau}$.

%It is interesting to note 
We note that for Poissonian resetting in $1d$, with no absorbing boundary,  the
steady-state position distribution, $p^*(x)= \sqrt{r/D}\;{\rm e}^{ -\sqrt{r/D} |x|}$ \cite{EM11a}, is proportional to the square root of the conjugate target distribution Eq.~\eqref{PTp}, when $|x|=R$, and the same  relation holds in $d>1$. This is reminiscent of the square-root principle for biased sampling which states that the best search distribution to sample in order to locate a target, is proportional
to the square root of the 
target distribution \cite{Snider11,Press09}. 
{\black Similar square-root relations have been noted in \cite{EM11b,KBGN17}.}
However,  the square-root relation between $P^*_T$ and $p^*$ is particular to Poissonian resetting and does not hold
for sharp restart or the heavy-tailed waiting time distribution.\\
\indent
It would be of  interest to determine conjugate target distributions for first passage problems
under resetting other than
%that are different from 
simple diffusion,  {\rm e.g.} diffusion in a potential \cite{RMR19,ANBND19,RRJCP1,RJCP,ARD22} or active Brownian motion \cite{EM18,KSB18,SBS20}.
{\black Also one could extend the range of resetting strategies to include
time-dependent waiting time distributions, where rates  depend either on the time since the last reset \cite{PKE16} or  on the absolute time since the process began \cite{KT19}. For example,  in the latter class a resetting rate $r(t) \propto 1/t$ can  reduce the MFPT, averaged over a Laplace distribution of targets, from the constant rate case.}
Importantly, several experimental groups have been able to implement
resetting protocols and reproduce theoretical results. In these experiments optical traps (with a finite width) are generally used to reset and confine colloidal particles and distribution of positions naturally emerge \cite{TPSRR20,BBPMC20,FBPCM21,BFPCM21}.  Thus, our theoretical predictions hold the promise of experimental realization.\\

%{\black As noted above, the target distribution is symmetric about the origin and there is no angular dependence.}
%{\black Here $Z = \int_0^\infty {\rm d}R\,  f^2(R)$ is the normalization constant chosen.
%, for convenience, so that
%$\int_0^\infty {\rm d}R \;S_d(R)\, P_T^*(R)=1$
%where $S_{d}(R)=
%    \frac{2 \pi^{d/2}}{\Gamma(d/2)} R^{d-1}$.}

\noindent {\it Acknowledgements}: We thank Richard Blythe and Satya Majumdar for reading and commenting on the manuscript. SR is thankful to the New Faculty Seed Grant by IISER Berhampur and INSPIRE Faculty (IFA19-CH326) Research Grant (DST/CHM/BPR/110124/063) by DST, Govt. of India. She gratefully acknowledges the Elizabeth Gardner Fellowship by the School of Physics \& Astronomy, University of Edinburgh that supported her during the initial stage of this work. For the purpose of open access, the authors have applied a Creative Commons Attribution (CC BY) licence to any Author Accepted Manuscript version arising from this submission.

%%%%%%%%%%%%%%%%%%%%%%%%%%%%%%%%%%%%%%%%%%%%%%%%%%%%%%%%%%%%%

%%%%%%%%%%%%%%%%%%%%%%%%%%%%%%%%%%%%%%%%%%%%%%%%%%%%%%%%%%%%%%

%%%%%%%%%%%%%%%%%%%%%%%%%%%%%%%%%%%%%%%%%%%%%%%%%%%%%%%%%%%%%%

\begin{widetext}
\section{\underline{Supplementary Material} }% Force line breaks with \\

\setcounter{equation}{0}
\setcounter{figure}{0}
\setcounter{table}{0}
\setcounter{page}{1}
\makeatletter
\renewcommand{\theequation}{S\arabic{equation}}
\renewcommand{\bibnumfmt}[1]{[S#1]}
\renewcommand{\citenumfont}[1]{S#1}
%\tableofcontents

%%%%%%%%%%%%%%%%%%%%%%%%%%%%%%%%%%%%%%%%%%%%%%%%%%%%%%%%%%%%
\section{\label{sec:level1} Derivation of Eq. (\ref{Tavp})}
Equation \eqref{Tavp} relates the mean first passage time to the target starting from distance $R$,  $\langle T_r(R) \rangle$, to the survival probability up to time $t$ starting from distance $R$, $Q_r(R,t)$, and the waiting time distribution for resetting, $\psi(t)$.
The first-passage time distribution for a dynamic process is given by the rate of decay of its survival probability. Therefore, the mean first passage time, $\langle T_r(R) \rangle$, satisfies the identity $\langle T_r\rangle=-\int_0^{\infty} {\rm d}t \;t (\partial Q_r(R,t)/\partial t) $ \cite{redner_S}. Integrating the rhs by parts one gets $\langle T_r\rangle =\int_0^{\infty}Q_r(R,t) {\rm d}t= [\widetilde Q_r(R,s)]_{s=0}$, where $\widetilde Q_r(R,s) = \int_0^\infty {\rm d} t\,{\rm e}^{-st}Q_r(R, t)$ is the Laplace transform of $Q_r(R, t)$. Following this relation, we put $s=0$ in Eq. \eqref{surv_LT} in the main text to obtain
\begin{equation}
     \langle T_r(R)\rangle = [\widetilde Q_r(R,s)]_{s=0} = \frac{ \int_0^\infty {\rm d}t\,\Psi(t) Q_0(R, t)}{1- \int_0^\infty {\rm d}t\, \psi(t)\,Q_0(R, t)}.
     \label{MFPT_s0}
\end{equation}
Since the waiting time distribution $\psi(t)$ is normalized, i.e., $\int_0^\infty \psi(t){\rm d}t = 1$, the denominator of 
Eq. \eqref{MFPT_s0} can be rewritten as 
\begin{equation}
\left[1- \int_0^\infty {\rm d}t\, \psi(t)\,Q_0(R, t)\right]= \int_0^\infty{\rm d}t \, \psi(t)[1- \,Q_0(R, t)].
\label{MFPT_den1}
\end{equation}
Integrating the numerator of Eq. \eqref{MFPT_s0} by parts we get 
\begin{equation}
\int_0^\infty {\rm d}t\,\Psi(t) Q_0(R, t) = \left[\left(\int_0^{t} {\rm d} t^{\prime} Q_0(R, t^{\prime})\right)\Psi(t)\right]_0^{\infty}-\int_0^{\infty}{\rm d} t\,\left[\frac{\partial \Psi(t)}{\partial t} \left(\int_0^{t}{\rm d}t^{\prime}\,Q_0(R,t^{\prime})\right)\right]. 
\label{MFPT_num1}
\end{equation}
For $t\to0$, the integral $\int_0^{t} {\rm d} t^{\prime} Q_0(R, t^{\prime})$ vanishes, and for $t\to \infty$, $\Psi(t)=1-\int_0^{t} {\rm d}t^{\prime}\psi (t^{\prime}) = 0$, as $\psi(t)$ is normalized. Therefore, the first term at the lhs of Eq. \eqref{MFPT_num1} is always zero. Moreover, since $\psi(t) = -\partial \Psi(t)/\partial t $, Eq. \eqref{MFPT_num1} reduces to 
\begin{equation}
\int_0^\infty {\rm d}t\,\Psi(t) Q_0(x, t) = \int_0^{\infty} {\rm d}t \;\psi(t)\int_0^{t}{\rm d}t^{\prime}\,Q_0(x,t^{\prime}).
\label{MFPT_num2}
\end{equation}
Substituting the numerator and denominator of  Eq. \eqref{MFPT_s0} by Eqs. \eqref{MFPT_num2} and \eqref{MFPT_den1}, respectively, we obtain Eq. (\ref{Tavp}) in the main text.

%%%%%%%%%%%%%%%%%%%%%%%%%%%%%%%%%%%%%%%%%%%%%%%%%%%%%%%%%%%
\section{\label{sec:level1} Derivation of Eq. (\ref{dTdpsi}) }

We consider $\langle T_r(R) \rangle$ as a functional of the waiting time distribution $\psi(t)$. The parameter $R$ is held fixed and  to lighten the notation we suppress the $R$ dependence and  write $\langle T_r[\psi(t)]\rangle$.
Following Eq. (\ref{Tav}) in the main text, we see that  
\begin{equation}
    \frac{\delta \overline{ \langle T_r[\psi(t)] \rangle}}{\delta \psi (t^{\prime})} = \int_0^\infty{\rm d}R\, P_T(R) \left[\frac{\delta\langle T_r[\psi(t)] \rangle}{\delta \psi (t^{\prime})}\right]\;.
    \label{Tav_target_fd}
\end{equation}
The functional derivative of $\langle T_r[\psi(t)] \rangle$ with respect to the waiting time distribution $\psi(t)$ is defined as
\begin{equation}
    \frac{\delta\langle T_r\rangle}{\delta \psi (t^{\prime})} = \lim_{\epsilon \to 0} \frac{\langle T_r[\psi(t) +\epsilon \delta (t-t^{\prime})] \rangle - \langle T_r[\psi(t)] \rangle}{\epsilon}\;.
    \label{Tav_fd_def}
\end{equation}
Utilizing Eq. (\ref{Tavp}) from the main text that shows the explicit dependence of $\langle T_r\rangle$ on $\psi(t)$, we obtain
\begin{eqnarray}
    \langle T_r[\psi(t) +\epsilon \delta (t-t^{\prime})] \rangle &=& \frac{\int_0^{\infty} {\rm d}t [\psi(t)+\epsilon \delta (t-t^{\prime})]\int_0^{t}{\rm d}t^{\prime}Q_0(R,t^{\prime})}{\int_0^{\infty}{\rm d}t[\psi(t)+\epsilon \delta (t-t^{\prime})][1- \,Q_0(R, t)]}\;,
    \label{Tav_fd_num1}
\end{eqnarray}
which, using the relation $\int_0^{\infty}{\rm d}t\; \delta(t-t^{\prime})f(t) = f(t^{\prime})$, can be simplified to 
\begin{eqnarray}
    \langle T_r[\psi(t) +\epsilon \delta (t-t^{\prime})] \rangle    
    &=& \frac{\int_0^{\infty} {\rm d}t\;\psi(t) \int_0^{t}{\rm d}t^{\prime} Q_0(R,t^{\prime}) +\epsilon  \int_0^{t^{\prime}}{\rm d}t \;Q_0(R,t)}{\int_0^{\infty}{\rm d}t\;\psi(t)[1- \,Q_0(R, t)]+\epsilon [1- \,Q_0(R, t^{\prime})]}\;.
    \label{Tav_fd_num2}
\end{eqnarray}
Further simplification of Eq. \eqref{Tav_fd_num2} leads to
\begin{eqnarray}
    \langle T_r[\psi(t) +\epsilon \delta (t-t^{\prime})] \rangle  
    &=& \frac{\int_0^{\infty} {\rm d}t\;\psi(t) \int_0^{t}{\rm d}t^{\prime} Q_0(R,t^{\prime}) +\epsilon  \int_0^{t^{\prime}}{\rm d}t \;Q_0(R,t)}{\int_0^{\infty}{\rm d}t\;\psi(t)[1- \,Q_0(R, t)]\left[1+\frac{\epsilon [1- \,Q_0(R, t^{\prime})]}{\int_0^{\infty}{\rm d}t\;\psi(t)[1- \,Q_0(R, t)]}\right]}\;,\nonumber\\
     &=& \frac{\int_0^{\infty} {\rm d}t\;\psi(t) \int_0^{t}{\rm d}t^{\prime} Q_0(R,t^{\prime}) +\epsilon  \int_0^{t^{\prime}}{\rm d}t \;Q_0(R,t)}{\int_0^{\infty}{\rm d}t\;\psi(t)[1- \,Q_0(R, t)]}\left[1-\frac{\epsilon [1- \,Q_0(R, t^{\prime})]}{\int_0^{\infty}{\rm d}t\;\psi(t)[1- \,Q_0(R, t)]}\right]\;\;(\mbox{since}\;\epsilon \;\mbox{is small}),\nonumber\\
     &=& \langle T_r[\psi(t)] \rangle + \epsilon \left[ \frac{\int_0^{t^{\prime}}{\rm d}t \;Q_0(R,t)}{\int_0^{\infty}{\rm d}t\;\psi(t)[1- \,Q_0(R, t)]}
-\frac{ [1- \,Q_0(R, t^{\prime})]\int_0^{\infty} {\rm d}t\;\psi(t) \int_0^{t}{\rm d}t^{\prime} Q_0(R,t^{\prime})}{\left(\int_0^{\infty}{\rm d}t\;\psi(t)[1- \,Q_0(R, t)]\right)^2}\right] + \mathcal{O}(\epsilon^2).\nonumber\\
    \label{Tav_fd_num3}
\end{eqnarray}
Plugging in the final (third) equality from Eq. \eqref{Tav_fd_num3} into Eq. \eqref{Tav_fd_def} and considering the limit $\epsilon \to 0$, we get an expression of $\frac{\delta\langle T_r\rangle}{\delta \psi (t^{\prime})}$. Putting that expression in Eq. (\ref{Tav}), one gets Eq.~(\ref{dTdpsi}). 
%%%%%%%%%%%%%%%%%%%%%%%%%%%%%%%%%%%%%%%%%%%%%%%%%%%%%%%%%%

\section{\label{sec:level1} Derivation of the LHS of Eq. (\ref{condlt})}

Incorporating the ansatz [introduced in Eqs.~(\ref{PT}, \ref{PT_f})] into Eq.~(\ref{dTdpsi}) we get  
\begin{multline}
    \frac{\delta \overline{\langle T_r \rangle}}{\delta \psi(t')}
    =Z^{-1} \int_0^\infty\hspace*{-0.2cm}
   {\rm d}R\, \left[
     \int_0^\infty \hspace{-0.2cm}{\rm d}t\, \psi(t)(1- \,Q_0(R, t))\int_0^{t'} {\rm d}\tau\,Q_0(R, \tau) - (1-Q_0(R,t'))\int_0^\infty \hspace*{-0.2cm}{\rm d}t\, \psi(t) \int_0^t \hspace*{-0.2cm}{\rm d}\tau\,Q_0(R, \tau)\right].
     \label{fd_PT_f}
\end{multline}
Laplace transforming Eq. \eqref{fd_PT_f} with respect to $t^{\prime}$ with Laplace variable $s$, we obtain
\begin{eqnarray}
    \mathcal{L}\left\{\frac{\delta \overline{\langle T_r \rangle}}{\delta \psi(t')}\right\}=\int_0^{\infty}{\rm d} t^{\prime} {\rm e}^{-s t^{\prime}}\left[\frac{\delta \overline{\langle T_r \rangle}}{\delta \psi(t')}\right]&=& \frac{\mathcal{A}(s)}{Z}
     \label{fd_PT_f_LT}
\end{eqnarray}
where
\begin{multline}
    \mathcal{A}(s) = 
    \int_0^{\infty}\hspace{-0.2cm}{\rm d} t^{\prime} {\rm e}^{-st^{\prime}}
    \left(\int_0^\infty\hspace{-0.2cm}{\rm d}R\; \left[
     \int_0^\infty \hspace{-0.2cm} {\rm d}t\, \psi(t)(1- \,Q_0(R, t))\int_0^{t'}\hspace{-0.2cm} {\rm d}\tau\,Q_0(R, \tau)
     - (1-Q_0(R,t'))
     \int_0^\infty \hspace{-0.25cm}{\rm d}t\, \psi(t) \int_0^t\hspace{-0.2cm}{\rm d}\tau\,Q_0(R, \tau)\right]\right).\\
     \label{A}
\end{multline}
Rearranging the order of integration over $R$ and $t^{\prime}$ of the rhs of Eq. \eqref{A2}, and then interchanging the integration variables $t$ and $t^{\prime}$ in the second term, we can rewrite it as
\begin{multline}
    \mathcal{A}(s) = 
    \int_0^\infty{\rm d}R\,\left\{\int_0^\infty {\rm d} t^{\prime} {\rm e}^{-s t^{\prime}} \int_0^{\infty} {\rm d}t\, \psi(t)(1- \,Q_0(R, t))\int_0^{t'} {\rm d}\tau\, Q_0(R, \tau)\right.\\
         - \int_0^\infty {\rm d} t\; {\rm e}^{-s t}(1-Q_0(R,t))
     \int_0^\infty \left.{\rm d}t^{\prime}\, \psi(t^{\prime}) \int_0^{t^{\prime}}{\rm d}\tau\,Q_0(R, \tau)\right\}.
     \label{A2}
\end{multline}
Rearranging the orders of integration, Eq. \eqref{A2} can be further simplified to
\begin{equation}
    \mathcal{A}(s) = 
    \int_0^{\infty} {\rm d}t\,\int_0^\infty {\rm d} t^{\prime} \left[{\rm e}^{-s t^{\prime}}  \psi(t)-{\rm e}^{-s t}\psi(t^{\prime}) \right]I(t,t'),
    \end{equation}
    where
    \begin{equation}
    I(t,t')=\int_0^{t^{\prime}}{\rm d}\tau\,\int_0^\infty {\rm d} R\, (1-Q_0(R,t))\; Q_0(R, \tau),
     \label{A3}
\end{equation}
which give Eqs. (\ref{condlt}) and (\ref{Idef}), respectively.

We now evaluate $I(t,t')$ in the case $d=1$ discussed in the main text, for which $S_d(R)=2$ and $Q_0(R,t) = \mbox{erf}( R/\sqrt{4Dt})$.
Then we have
\begin{eqnarray}
    I(t,t')&=& 2\int_0^{t^{\prime}}{\rm d}\tau\,\int_0^\infty {\rm d}R\, \mbox{erfc}(R/\sqrt{4Dt})\, \mbox{erf}( R/\sqrt{4D\tau})\\
    &=& \frac{4 D^{1/2}}{\pi^{1/2}}\int_0^{t'}{\rm d}\tau\,\left[ (\tau+t)^{1/2}-\tau^{1/2} \right]\\
    &=& \frac{8}{3}\frac{D^{1/2}}{\pi^{1/2}}\left[ (t'+t)^{3/2} -t^{3/2}-(t')^{3/2}\right],
    \end{eqnarray}
where we have used identity 4.7.1 of \cite{NG1969_S}
\begin{equation}
    \int_0^\infty \mbox{\hspace{-0.3cm}}{\rm d} R\,\, \mbox{erfc}(a^{-1/2}R) \mbox{erf}(b^{-1/2}R)\,
    = \frac{1}{\sqrt{\pi}}\left( (a+b)^{1/2}-b^{1/2}\right)\label{erfidents}
\end{equation}

In arbitrary dimension $d$, the proof  that $I(t,t')$ is symmetric, requires taking the  double Laplace transform of $I(t,t')$ with respect to $t$ and $t'$ and using identities for integrals of modified Bessel functions. The details will be presented elsewhere.

%%%%%%%%%%%%%%%%%%%%%%%%%%%%%%%%%%%%%%%%%%%%%%%%%%%%%%%%%%

\section{Condition for $\psi(\tau)$ to be normalizable}
The normalization constant $Z$ in the case $d=1$ is given by 
%Equation ? of the main text
\begin{eqnarray}
    Z&=& 2\int_0^\infty{\rm d}R\, f^2(R) \nonumber \\
    &=& 2\int_0^\infty{\rm d}R \int_0^\infty {\rm d}t\, \psi(t)\int_0^\infty {\rm d}t'\, \psi(t')
    \mbox{erfc}(R/\sqrt{4Dt})\, \mbox{erfc}( R/\sqrt{4Dt'})\nonumber\\
    &=& \frac{4 D^{1/2}}{\pi^{1/2}}
    \int_0^\infty {\rm d}t\, \psi(t)\int_0^\infty {\rm d}t' \psi(t')\left[t^{1/2}+(t')^{1/2}-(t+t')^{1/2}\right],\label{Zs1}
\end{eqnarray}
where we have used  the definitions
\begin{equation}
    f(R) = \int_0^\infty {\rm d}t\, \psi(t) (1-Q_0(R,t))\;,\label{fs}
\end{equation}
\begin{equation}
    Q_0(R,t) = \mbox{erf}(R/\sqrt{4Dt})\;,\label{Q0s}
\end{equation}
and identity 4.7.2  from \cite{NG1969_S}
\begin{equation}
    \int_0^{\infty}{\rm d} R\,\mbox{erfc}(a^{-1/2}R)\,\mbox{erfc}(b^{-1/2}R) =\frac{1}{\sqrt \pi}(a^{1/2}+ b^{1/2} -(a+b)^{1/2})\;.
\end{equation}
We  let $t' = ut$,  then \eqref{Zs1} becomes
\begin{equation}
Z=  \frac{4 D^{1/2}}{\pi^{1/2}}
    \int_0^\infty {\rm d}t\, \psi(t)t^{3/2}\int_1^\infty {\rm d}u\, \psi(ut)\left[1+u^{1/2}-(1+u)^{1/2}\right].\label{Zs2}
\end{equation}
We now assume that the large $t$ behaviour of the waiting time distribution is
\begin{equation}
    \psi(t) \sim t^{-\alpha}.
\end{equation}
Then to determine whether integral over $u$ in \eqref{Zs2} converges, we expand the integrand for large $u$
\begin{equation}
    \int_1^\infty {\rm d}u\, \psi(ut)\left[1+u^{1/2}-(1+u)^{1/2}\right]
    \sim\frac{1}{t^\alpha} \int^\infty {\rm d}u\, \frac{1}{ u^\alpha}
    \left[ 1-O(u^{-1/2})\right]
\end{equation}
which converges if $\alpha>1$.
Similarly, to determine when the $t$ integral in \eqref{Zs2} converges, we consider the large $t$ behaviour of the integrand. We require the following integral to converge
\begin{equation}
    \int^\infty {\rm d}t\, t^{3/2} t^{-2\alpha}
\end{equation}
which requires $\alpha > 5/4$.

Thus we conclude  that if $\psi(t)$ decays faster than $t^{-5/4}$, $Z$ will be finite. In other words, for  the conjugate of $\psi(t)$, Eq. \eqref{PT}, to be normalizable we require  
the expectation value of $t^{1/4}$, where $t$ is the random waiting time,
\begin{equation}
   \langle t^{1/4}\rangle =  \int_0^\infty {\rm d}t\, \psi(t) t^{1/4}\;, 
\end{equation} to be finite.

\section{Condition for MFPT  averaged over conjugate  target distribution to be finite}
From the main text the mean first passage time averaged over the conjugate target distribution $P^*_T(R)$ is
\begin{eqnarray}
    \overline{ \langle T_r \rangle}&=& 2\int_0^\infty{\rm d}R\,P_T^*(R)
    \left[ \frac{\int_0^\infty{\rm d}t\,\psi(t)\int_0^t dt'' Q_0(R,t'')}{f(R)}\right]\\
    &=& \frac{2}{Z} \int_0^\infty{\rm d}R\, f(R)\int_0^\infty {\rm d}t\, \psi(t)\int_0^t {\rm d}t''\, Q_0(R,t'')\\
    &=& \frac{2}{Z} \int_0^\infty {\rm d}t'\,\psi(t')\int_0^\infty {\rm d}t\,\psi(t)\int_0^t {\rm d}t''\,\int_0^\infty{\rm d}R\, (1-Q_0(R,t')) Q_0(R,t'')\;,\\
     &=& \frac{2}{Z} \int_0^\infty {\rm d}t'\,\psi(t')\int_0^\infty {\rm d}t\,\psi(t)\int_0^t {\rm d}t''\,\int_0^\infty{\rm d}R\, \mbox{erfc}(R/\sqrt{4Dt'}) \mbox{erf}(R/\sqrt{4Dt'})\;,
\end{eqnarray}
where we have used \eqref{fs} and \eqref{Q0s}.
We now use identity \eqref{erfidents}
to perform the $R$ integral and obtain
\begin{eqnarray}
    \overline{ \langle T_r \rangle}
    &=& \frac{2}{Z}\int_0^\infty {\rm d}t\,\psi(t)\int_0^\infty {\rm d}t'\,\psi(t')\int_0^t {\rm d}t''\left(\frac{4D}{\pi}\right)^{1/2}
    \left[ (t'+t'')^{1/2}-t''^{1/2}\right]\\
    &=& \frac{4}{3Z}\left(\frac{4D}{\pi}\right)^{1/2}  \int_0^\infty {\rm d}t\,\psi(t)\int_0^\infty {\rm d}t'\,\psi(t')
    \left[ (t'+t)^{3/2}-(t')^{3/2}-t^{3/2}\right]\;.\label{Tavs}
\end{eqnarray}
We  let $t' = ut$,  then \eqref{Tavs} becomes
\begin{equation}
 \overline{ \langle T_r \rangle}=  \frac{4}{3Z}\left(\frac{4D}{\pi}\right)^{1/2} 
    \int_0^\infty {\rm d}t\, \psi(t)t^{5/2}\int_1^\infty {\rm d}u\, \psi(ut)\left[(1+u)^{3/2}-1-u^{3/2}\right].\label{Tavs2}
\end{equation}
We now assume that the large $t$ behaviour of the waiting time distribution is
\begin{equation}
    \psi(t) \sim t^{-\alpha}.
\end{equation}
Then to determine whether integral over $u$ in \eqref{Tavs2} converges we expand the integrand for large $u$
\begin{equation}
    \int_1^\infty {\rm d}u\, \psi(ut)  \left[(1+u)^{3/2}-1-u^{3/2}\right]
    \sim\frac{1}{t^\alpha} \int^\infty {\rm d}u\, \frac{1}{u^\alpha}\left[\frac{3}{2}u^{1/2}
    +O(u^{-1/2})\right]
\end{equation}
which converges if $\alpha>3/2$.
Similarly, to determine when the $t$ integral in \eqref{Zs2} converges we consider the large $t$ behaviour of the integrand. We require the following integral to converge
\begin{equation}
    \int^\infty {\rm d}t\, t^{5/2} t^{-2\alpha}
\end{equation}
which requires $\alpha > 7/4$.

Thus we conclude  that if $\psi(t)$ decays faster than $t^{-7/4}$, $\overline{\langle T_r \rangle}$ will be finite. In other words we require 
\begin{equation}
   \langle t^{3/4}\rangle =  \int_0^\infty {\rm d}t\, \psi(t) t^{3/4} 
\end{equation} to be finite.

%%%%%%%%%%%%%%%%%%%%%%%%%%%%%%%%%%%%%%%%%%%%%%%%%%%%
\section{Sharp restart is optimal for fixed target}

As noted in the Introduction, sharp reset has been of interest as  it minimizes the MFPT when the distance $R_0$ to the target is known.
A fixed target at distance $R_0$ from the resetting site (taken to be the origin)  corresponds to a target distribution $P_T (R) = \delta(R-R_0)$. It has been shown that for this scenario sharp restart minimizes the mean first passage time when  the deterministic period of resetting, $\tau$, is suitably chosen \cite{PR17_S,CS18_S}.
Here we show how this result is recovered within our formalism.

To see this
we insert $\psi(t)=\delta(t-\tau)$ and $P_T(R) = \delta(R-R_0)$ in Eq. \eqref{dTdpsi} yielding
\begin{equation}
    \frac{\delta \overline{\langle T_r \rangle}}{\delta \psi(t')}
    = 
   \frac{1-Q_0(R_0,t')}
    {1- \,Q_0(R_0, \tau)}\left[ G(t')-G(\tau) \right],
    \label{dTdpsis}
\end{equation}
where $G(t') = \int_0^{t'} {\rm d}t\,Q_0(R_0, t)/(1- \,Q_0(R_0, t^{\prime}))$.
If $\tau$ is chosen as the value of $t'$ that minimizes  $G(t')$ \cite{PR17_S,CS18_S}, we have
\begin{equation}
    \frac{\delta \overline{\langle T_r \rangle}}{\delta \psi(t')} \geq 0\;,\label{varsr}
\end{equation}
with the equality only holding when $t' = \tau$. Then sharp reset with period $\tau$ minimizes $\overline{\langle T_r \rangle}$ because of the constraints that $\psi(t) \geq 0$ and $\int_0^\infty dt\,\psi(t) =1$, i.e.,  variation of the delta distribution  implies increasing $\psi(t')$ when $t'\neq \tau$ and 
decreasing $\psi(\tau)$.
Consequently, since Eq. \eqref{varsr} holds,  $\overline{\langle T_r \rangle}$ always increases.

%%%%%%%%%%%%%%%%%%%%%%%%%%%%%%%%%%%%

\section{Derivation of conjugate target distrbution for Poissonian resetting in arbitrary dimension Eq. (\ref{PTd})}

For $d>1$ we consider the target as an absorbing sphere
of radius $a$ centred at  $\vec x$ (where $|\vec{x}|=R  > a$). 
The particle starts at the origin  and undergoes diffusion with diffusion constant $D$ and stochastic resetting to the origin. 
When the particle reaches the surface of the target sphere, it is absorbed.

For general $d$ the Laplace transform of the survival probability without resetting  is
known in closed form \cite{redner_S,BMS13_S}
\begin{equation}
    \tilde Q_0(R,s) = \frac{1}{s} \left[ 1 - \left(\frac{R}{a}\right)^\nu \frac{K_\nu(R(s/D)^{1/2}}{K_\nu(a(s/D)^{1/2}}\right]\label{survLT},
\end{equation}%
where $\nu = 1-d/2$ and $K_\nu(z)$ is a modified Bessel function of the second kind. 
The proof of Eqs.~(\ref{PT}, \ref{PT_f}) follows the one-dimensional case, but we now take the double Laplace transform
$\tilde I(u,v) = \int_0^\infty {\rm d}t\,{\rm e}^{-ut}\int_0^\infty {\rm d}t'\,{\rm e}^{-vt'}I(t,t')$ %and require that it
that is symmetric under  $u \leftrightarrow v$.
This can be shown  using  integral identities for  the product of functions $K_\nu(z)$ \cite{GR_S}. 
Thus in any  dimension $d$,
Eq. \eqref{PT} gives the particular target 
distribution for which an arbitrary $\psi(t)$ 
is extremal.

The waiting time distribution for Poissonian resetting is given by $\psi(t) = r {\rm e}^{-r t}$. Plugging  that into the definition of $f(R)$ (given by Eq.~\eqref{PT_f} in the main text), we get $f(R)=r \int_0^{\infty}{\rm d}t\; {\rm e}^{-r t} (1-Q_0(R,t))$. The first term of the rhs is simply $\int_0^{\infty}{\rm d}t\,r\,{\rm e}^{-r t}=1$ and the second term can be written as a Laplace transform of $Q_0(R,t)$, leading to $f(R)=1-r\,\tilde{Q_0}(R,r)$. Utilizing Eq. (\ref{survLT}), we thus obtain 
\begin{equation}
f(R) = \left(\frac{R}{a}\right)^\nu \frac{K_\nu(R(r/D)^{1/2}}{K_\nu(a(r/D)^{1/2}}.
\label{f_d}
\end{equation}
We then insert this expression for $f(R)$ into
\begin{equation}
    P_T^*(R) =   Z^{-1}f^2(R),\label{PT_S} 
\end{equation}
where
\begin{eqnarray}
 Z &=& \int_0^\infty {\rm d}R\,S_d(R)  f^2(R) \\
\mbox{and}\quad  S_{d}(R)&=&
    \frac{\pi^{d/2}}{\Gamma(d/2)} R^{d-1}
\end{eqnarray}
which leads to
\begin{equation}
    P_T^*(R) = Z^{-1} \left[ \left(\frac{R}{a}\right)^\nu \frac{K_\nu(R(r/D)^{1/2}}{K_\nu(a(r/D)^{1/2}}\right]^2 .
    \label{PTd_S}
\end{equation}
%where $\nu = 1-d/2$ and $K_\nu(z)$ is a modified Bessel function of the second.
This distribution is normalizable for $d<4$.
In 1$d$ one can check using 
\begin{equation}
 K_{1/2}(z) = (\pi/2z)^{1/2} {\rm e}^{-z}
 \end{equation} 
that the target distribution reduces to 
\begin{equation}
P_T^*(R) = \sqrt{\frac{r}{D}} {\rm e}^{-2 \sqrt{\frac{r}{D}} R}\;, 
\end{equation}
as given by Eq.~(\ref{PTp}) in the main text.

%%%%%%%%%% Prefix a "S" to all equations, figures, tables and reset the counter %%%%%%%%%%
%\tableofcontents

\end{widetext}

\begin{thebibliography}{50}
\vspace{0.2cm}

\bibitem{EMS20} M. R. Evans, S. N. Majumdar, and G. Schehr, Stochastic resetting and applications, J. Phys. A: Math. Theor. \textbf{53}, 193001 (2020).

\bibitem{LSZ93} M. Luby, A. Sinclair, and D. Zuckerman, Optimal speedup of las vegas algorithms, Information Processing Letters \textbf{47}, 173 (1993).

\bibitem{BRH22} O. Blumer, S. Reuveni, and B. Hirshberg, Stochastic resetting for enhanced sampling, J. Chem. Phys. Lett. {\bf 13}, 11230–11236 (2022).

\bibitem{BRH24} O. Blumer, S. Reuveni, and B. Hirshberg, Combining stochastic resetting with metadynamics to speedup molecular dynamics simulations, Nature Communications {\bf 15}, 240 (2024).

\bibitem{RUK14} S. Reuveni, M. Urbakh, and J. Klafter, Role of substrate unbinding in Michaelis–Menten enzymatic reactions, Proc. Natl. Acad. Sci. USA {\bf 111}, 4391 (2014).

\bibitem{RRU15} T. Rotbart, S. Reuveni, and M. Urbakh, Michaelis-Menten reaction scheme as a unified approach towards
the optimal restart problem, Phys. Rev. E {\bf 92}, 060101
(2015).

\bibitem{BPMR23} A. Biswas, A. Pal, D. Mondal, and S. Ray, Rate enhancement of gated drift-diffusion process by optimal resetting, J. Chem. Phys. {\bf 159}, 054111 (2023).

\bibitem{BS14} D. Boyer and C. Solis-Salas, Random walks with preferential relocations to places visited in the past and their application to biology, Phys. Rev. Lett. {\bf 112}, 240601 (2014).

\bibitem{VCMLNA22} O. Vilk, D. Campos, V. M\'endez, E. Lourie, R. Nathan, and M. Assaf, Phase transition in a non-Markovian animal exploration model with preferential returns, Phys.
Rev. Lett. {\bf 128}, 148301 (2022).

\bibitem{EMS22} M. R. Evans, S. N. Majumdar, and G. Schehr, An exactly solvable predator prey model with resetting, J. Phys. A: Math. Theor. {\bf 55}, 274005 (2022).

\bibitem{RLSG16} \'E. Rold\'an, A. Lisica, D. S\'anchez-Taltavull, and S.W. Grill, Stochastic resetting in backtrack recovery by RNA polymerases,
Phys. Rev. E {\bf 93}, 062411 (2016).

\bibitem{Brf2020a}  P. C. Bressloff, Modeling active cellular transport as a directed search process with stochastic resetting and delays, J. Phys. A: Math. Theor. {\bf 53}, 355001 (2020). 

\bibitem{CM15} D. Campos and V. M\'endez, Phase transitions in optimal search times: How random walkers should combine resetting and flight scales, Phys. Rev. E {\bf 92}, 062115 (2015).

\bibitem{BDBR16} U. Bhat, C. De Bacco, and S. Redner, Stochastic search with poisson and deterministic resetting, J. Stat. Mech.: Theor. Expt. {\bf 2016}, 083401 (2016).
\bibitem{PR17} A. Pal and S. Reuveni, First passage under restart, Phys. Rev. Lett. {\bf 118}, 030603 (2017).
\bibitem{CS18} A. Chechkin and I. M. Sokolov, Random Search with Resetting: A Unified Renewal Approach, Phys. Rev. Lett. {\bf 121}, 050601 (2018).

\bibitem{B18} S. Belan, Restart Could Optimize the Probability of Success in a Bernoulli Trial, Phys. Rev. Lett. {\bf 120}, 080601 (2018).

\bibitem{BMS23}
M. Biroli, S. N. Majumdar, G. Schehr, Critical number of walkers for diffusive search processes with resetting, Phys. Rev. E,  {\bf 107}, 064141 (2023).

\bibitem{SBEM24}
J. C. Sunil, R. A. Blythe, M. R. Evans, S. N. Majumdar, Minimizing the Profligacy of Searches with Reset, Phys. Rev. E, {\bf 110} 054122 (2024).


\bibitem{EM11a} M. R. Evans and S. N. Majumdar, Diffusion with stochastic resetting, Phys. Rev. Lett. {\bf 106}, 160601 (2011).

\bibitem{EM11b} M. R. Evans and S. N. Majumdar, Diffusion with optimal resetting, J. Phys. A: Math. Theor. {\bf 44}, 435001 (2011).



\bibitem{WEM13} J. Whitehouse, M. R. Evans, and S. N. Majumdar, Effect of partial absorption on diffusion with resetting, Phys. Rev. E {\bf 87}, 022118 (2013).

\bibitem{EM14} M. R. Evans and S. N. Majumdar, Diffusion with resetting in arbitrary spatial dimension, J. Phys. A: Math. Theor. {\bf 47}, 285001 (2014).

\bibitem{MSS15} S. N. Majumdar, S. Sabhapandit, and G. Schehr, Dynamical transition in the temporal relaxation of stochas-
tic processes under resetting, Phys. Rev. E {\bf 91}, 052131 (2015).


\bibitem{CS15} C. Christou, and A. Schadschneider, Diffusion with resetting in bounded domains, J. Phys. A: Math. Theor. {\bf 48}, 285003 (2015).

\bibitem{KGN15} {\black {\L}. Ku\'smierz, and E. Gudowska-Nowak, Optimal first-arrival times in L\'evy flights with resetting, Phys. Rev. E {\bf 92}, 052127 (2015). }
\bibitem{KT19} {\black {\L}. Ku\'smierz, and T. Toyoizumi, Robust random search with scale-free stochastic resetting, Phys. Rev. E {\bf 100}, 032110 (2019). }


\bibitem{EM16} S. Eule and J. J. Metzger, Non-equilibrium steady states of stochastic processes with intermittent resetting, New J. Phys. {\bf 18}, 033006 (2016).

\bibitem{PKE16} A. Pal, A. Kundu, and M. R. Evans, Diffusion under time-dependent resetting, J. Phys. A: Math. Theor. {\bf 49}, 225001 (2016).

\bibitem{NG16} A. Nagar and S. Gupta, Diffusion with stochastic resetting at power-law times, Phys. Rev. E {\bf 93}, 060102(R) (2016).



\bibitem{G19} D. Gupta, Stochastic resetting in underdamped Brownian motion, J. Stat. Mech: Theor. Expt. {\bf 2019}, 033212 (2019). 

\bibitem{BS20a} A. Bodrova and I. M. Sokolov, Resetting processes with non-instantaneous return, Phys. Rev. E {\bf 101}, 052130 (2020).

\bibitem{BRR20} B. De Bruyne, J. Randon-Furling, and S. Redner, Optimization in First-Passage Resetting, Phys. Rev. Lett. {\bf 125}, 050602 (2020).

\bibitem{MMSC21}  V. M\'endez, A. Mas\'o-Puigdellosas, T. Sandev, and D. Campos, Continuous time random walks under Markovian resetting, Phys. Rev. E {\bf 103}, 022103 (2021).


\bibitem{DBMS22} B. De Bruyne, S. N. Majumdar, and G. Schehr, Resetting Brownian Bridges via Enhanced Fluctuations, Phys. Rev. Lett. {\bf 128}, 200603 (2022).

\bibitem{S23}  I. M. Sokolov, Linear Response and Fluctuation-Dissipation Relations for Brownian Motion under Resetting, Phys. Rev. Lett. {\bf 130}, 067101 (2023).

\bibitem{DBM23}  B. De Bruyne and F. Mori, Resetting in stochastic optimal control, Phys. Rev. Research {\bf 5}, 013122 (2023). 

\bibitem{MOK23}  F. Mori, K. S. Olsen, and S. Krishnamurthy, Entropy production of resetting processes, Phys. Rev. Research {\bf 5}, 023103  (2023). 

\bibitem{SBEM23}
J. C. Sunil, R.A. Blythe, M.R. Evans, S.N. Majumdar,
The cost of stochastic resetting,
J. Phys. A: Math. Theor. {\bf 56}, 395001 (2023)

\bibitem{BFM23} E. Barkai, R. Flaquer-Galm\'es, and V. M\'endez, Ergodic properties of Brownian motion under stochastic resetting, Phys. Rev. E {\bf 108}, 064102 (2023). 

\bibitem{JBPD} S. Jain, D. Boyer, A. Pal, and L. Dagdug, Fick–jacobs description and first passage dynamics for diffusion in a channel under stochastic resetting, J. Chem. Phys. {\bf 158}, 054113 (2023).

\bibitem{ER20} I. Eliazar and S. Reuveni, Mean-performance of sharp restart I: statistical roadmap, J. Phys. A: Math. Theor. {\bf 53}, 405004 (2020).

\bibitem{ER21} I. Eliazar and S. Reuveni, Mean-performance of sharp restart II: inequality roadmap, J. Phys. A: Math. Theor. {\bf 54}, 355001 (2021).

\bibitem{TPSRR20}
O. Tal-Friedman, A. Pal, A. Sekhon, S. Reuveni, Y. Roichman,
Experimental realization of diffusion with stochastic resetting,
J. Phys. Chem. Lett. {\bf 11}, 7350 (2020).

\bibitem{BBPMC20} B. Besga, A. Bovon, A Petrosyan, S. N. Majumdar,  
S. Ciliberto, Optimal mean first-passage time for a Brownian searcher subjected to resetting: experimental and theoretical results. Phys. Rev. Res. {\bf 2}, 032029 (2020).

\bibitem{FBPCM21}
F. Faisant, B. Besga, A. Petrosyan, S. Ciliberto, S. N. Majumdar,
Optimal mean first-passage time of a Brownian searcher with resetting in one and two dimensions: experiments, theory and numerical tests,
J. Stat. Mech.: Theory Expt. {\bf 2021}, 113203 (2021).

\bibitem{BFPCM21}
B. Besga, F. Faisant, A. Petrosyan, S. Ciliberto, and S. N. Majumdar,  Dynamical phase transition in the first-passage probability of a Brownian motion. Physical Review E, {\bf 104}, L012102 (2021).

{\black
\bibitem{PKR20} A. Pal, {\L}. Ku\'smierz, and S. Reuveni, Search with home returns provides advantage under high uncertainty, Phys. Rev. Research {\bf 2}(4), 043174 (2020).

\bibitem{ABGMRTRR24}

A Altshuler, OL Bonomo, N Gorohovsky, S Marchini, E Rosen, O Tal-Friedman, S Reuveni, and Y Roichman,
Environmental memory facilitates search with home returns,
Phys. Rev. Research {\bf 6} (2), 023255 (2024)



\bibitem{Bell} W. J. Bell, Searching Behaviour: The Behavioural Ecology of Finding Resources, Chapman \& Hall Animal Behaviour Series, Springer Dordrecht (1990).

\bibitem{BLMV11} O. B\'enichou, C. Loverdo, M. Moreau, and R. Voituriez, Intermittent search strategies, Rev. Mod. Phys. {\bf 83}, 81 (2011).


\bibitem{TVB12} V. Tejedor, R. Voituriez, and O. B\'enichou, 
Optimizing persistent random searches, Phys. Rev. Lett. {\bf 108}, 088103 (2012). }



\bibitem{Redner} S. Redner, A Guide to First-Passage Processes, Cambridge University Press, (2001).

\bibitem{BMS13}
A. J. Bray, S. N. Majumdar, and G. Schehr, Persistence and first-passage properties in non-equilibrium systems, Adv. Phys. {\bf 62}, 225 (2013).

\bibitem{ER24-SM}  See Supplemental Material for full derivations of Equations \eqref{Tavp}, \eqref{dTdpsi}, \eqref{condlt} and \eqref{PTd}; the condition for $\psi(\tau)$ to be normalizable, and the case of a delta function target distribution.

\bibitem{KBGN17} {\black {\L}. Ku\'smierz, M. Bier, and E. Gudowska-Nowak, Optimal potentials for diffusive search
strategies, J. Phys. A: Math. Theor. {\bf 50}, 185003 (2017). }
\bibitem{NG1969} E. W. Ng and M. Geller, A table of integrals of the error functions, Journal of Research of the National Bureau of Standards B {\bf 73}, 1 (1969).


\bibitem{Snider11} J. Snider, Optimal random search for a single hidden target, Phys. Rev. E {\bf 83}, 011105 (2011).

\bibitem{Press09} W. H. Press, Strong profiling is not mathematically optimal for discovering rare malfeasors, Proc. Natl. Acad. Sci. USA {\bf 106}, 1716 (2009).

\bibitem{RMR19} S. Ray, D. Mondal, and S. Reuveni, P\'eclet number governs transition to acceleratory restart in drift-diffusion, J. Phys. A: Math. Theor. {\bf 52}, 255002 (2019).

\bibitem{ANBND19} S. Ahmad, I. Nayak, A. Bansal, A. Nandi, D. Das, First passage of a particle in a potential under stochastic resetting: A vanishing transition of optimal resetting rate, 
Phys. Rev. E {\bf 99}, 022130 (2019). 


\bibitem{RRJCP1} S. Ray and S. Reuveni, Diffusion with resetting in a logarithmic potential, J. Chem. Phys. {\bf 152}, 234110 (2020).

\bibitem{RJCP} S. Ray, Space-dependent diffusion with stochastic resetting: A first-passage study, The Journal of Chemical Physics, {\bf 153}, 234904 (2020).

\bibitem{ARD22}  S. Ahmad, K. Rijal, and D. Das, First passage in the presence of stochastic resetting and a potential barrier, Phys. Rev. E {\bf 105}, 044134 (2022).


\bibitem{EM18}
M. R. Evans and S. N. Majumdar, Run and tumble particle under resetting: a renewal approach, J. Phys. A: Mat.  Theor. {\bf 51}, 475003 (2018).

\bibitem{KSB18} V Kumar, O Sadekar, U Basu, Active Brownian motion in two dimensions under stochastic resetting, Phys. Rev. E {\bf 102}, 052129 (2018). 


\bibitem{SBS20} I. Santra, U. Basu, S. Sabhapandit, Run-and-tumble particles in two dimensions under stochastic resetting conditions, Phys. Rev. E {\bf 101}, 062120 (2020). 

\bibitem{GR} I. S. Gradshteyn and I. M. Ryzhik, Table of integrals, series, and products, Academic press, (1980). Cited in \cite{ER24-SM}.



%\bibitem{PCM14} {\black Palyulin, V. V., Chechkin, A. V.,  Metzler, R. (2014). L\'evy flights do not always optimize random blind search for sparse targets. Proceedings of the National Academy of Sciences, {\bf 111}, 2931-2936.}

\end{thebibliography}

\vspace{0.5cm}

\begin{thebibliography}{10}
\bibitem{redner_S} S. Redner, A Guide to First-Passage Processes, Cambridge University Press, (2001).

\bibitem{NG1969_S} E. W. Ng and M. Geller, A table of integrals of the error functions, Journal of Research of the National Bureau of Standards B {\bf 73}, 1 (1969).

\bibitem{PR17_S} A. Pal and S. Reuveni, First passage under restart, Phys. Rev. Lett. {\bf 118}, 030603 (2017).
\bibitem{CS18_S} A. Chechkin and I. M. Sokolov, Random Search with Resetting: A Unified Renewal Approach, Phys. Rev. Lett. {\bf 121}, 050601 (2018).

\bibitem{BMS13_S}
A. J. Bray, S. N. Majumdar, and G. Schehr, Persistence and first-passage properties in non-equilibrium systems, Adv. Phys. {\bf 62}, 225 (2013).


\bibitem{GR_S} I. S. Gradshteyn and I. M. Ryzhik, Table of integrals, series, and products, Academic press, (1980).
\end{thebibliography}
\end{document}